\documentclass[useAMS,graphicx,usenatbib]{biom}
\usepackage{amsmath,amssymb}
\usepackage{hyperref}
\usepackage{booktabs}
\usepackage{graphicx}
\usepackage{caption}
\usepackage{subcaption}
\usepackage{booktabs}
\usepackage{ulem}
\usepackage{parskip}
\usepackage{bm}
\usepackage{enumitem}

\def\bSig\mathbf{\Sigma}

\title[Introducing precision-weighted bias]{Introducing precision-weighted bias as a performance measure to inform the inclusion of adaptive designs in meta-analysis}

\author{Martin Law$^{1,2*}$\email{martin.law@mrc-bsu.cam.ac.uk}, 
David S.\ Robertson$^{1}$,
Sofía S.\ Villar$^{1}$,
Tim P.\ Morris${^3}$,\\
Babak Choodari-Oskooei${^4}$,
Thomas Jaki$^{1,5}$,
and Ian R.\ White${^4}$\\
$^{1}$Medical Research Council Biostatistics Unit, University of Cambridge, Cambridge CB2 0SR, UK \\
$^{2}$Royal Papworth Hospital, Cambridge, UK\\
$^{3}$Statistical Methodology, Novartis Pharmaceuticals UK Ltd.\\
${^4}$UCL Innovative Clinical Trials Unit, University College London, UK\\
${^5}$Department of Machine Learning and Statistics, University of Regensburg, DE
}

\begin{document}

\begin{abstract}
  We propose a novel, intuitive measure of statistical performance: precision-weighted bias. Precision-weighted bias is defined as the unconditional bias of an estimator weighted by the degree of information (precision) it contains. Current guidelines, such as \textsc{grade} and \textsc{consort}, often view the potential for increased bias in adaptive designs as a deterrent for the inclusion of such designs in systematic reviews. However, we demonstrate that the bias in a common-effect meta-analysis is approximately equal to the precision-weighted average of the precision-weighted biases of its constituent studies, rather than of their unweighted unconditional biases.  Through simulation studies, we show that while adaptive designs may exhibit unweighted bias, they frequently have zero precision-weighted bias. Consequently, including these designs often results in a negligible change to the overall meta-analysis bias. These results suggest that precision-weighted bias is a superior indicator for determining whether to include an adaptive design in a meta-analysis. We recommend that precision-weighted bias be used as a standard complement to unweighted unconditional and conditional bias in simulation studies to support more inclusive and accurate evidence synthesis.
\end{abstract}


\maketitle

\section{Introduction}

Statistical techniques are often evaluated by running simulation studies, where statistical methods are applied to the data sets and their performance examined. A central measure of performance is (statistical) bias, defined as a systematic tendency for an estimate to differ from its true value. In the context of clinical trials, a simulation study involves creating computer-generated data sets representing trial data that could have been obtained when using a specific study design and under certain assumptions~\citep{lee2026OCTAVE}. Using simulation studies, we can estimate for a particular clinical trial design the bias of a treatment effect estimator, $\text{Bias} = E[\hat{\theta}]-\theta$, where $\theta$ and $\hat{\theta}$ denote the true treatment effect and an estimator, respectively.

Adaptive designs are trial designs where some aspect of the ongoing trial is modified, typically in a pre-specified way, based on interim information~\citep{ICHE20,pallmann2018adaptive}. Simulation studies are often used to evaluate statistical bias in adaptive designs: simulation may be the only way to obtain performance measures such as bias. Importantly, the realised adaptations can influence bias. For example, in adaptive design trials that stopped early (only) for benefit, the treatment effect size is overestimated~\citep[e.g.][]{Whitehead1986Dec}. This bias is conditional on the fact that the trial stopped early. In contrast to unconditional bias, defined above simply as ``bias'', conditional bias (in general) is the difference between the expectation of an estimator given a sample statistic $s$, such as whether or not the trial stopped early, and the true value of the estimand, defined as $\text{conditional bias} = E[\hat{\theta}|s]-\theta$. In certain circumstances, conditional bias may be of little or no concern~\citep{marschner2021general}. In general, when the data from a single adaptive design trial are to be analysed on their own, we may be more interested in the conditional bias; when a trial is just one of a collection of trials, for example, in a meta-analysis, we may be more interested in the unconditional bias. \cite{robertson2021point} have reviewed various methods to adjust for conditional and unconditional bias. These methods may increase the variance of estimates.

Some existing literature regards adaptive designs' potential for increased bias as a problem, particularly with respect to meta-analysis or systematic review. An example is \textsc{grade} (Grading of Recommendations Assessment, Development, and Evaluation), a framework for summarising quality of evidence in a study~\citep{GRADE_site}. ~\cite{guyatt2011grade}, as part of \textsc{grade}, recommend that systematic reviews should conduct a sensitivity analysis excluding trials that stopped early for benefit, and then consider the sensitivity analysis as more credible if the results disagree with the analysis that includes such trials. 

The \textsc{consort} (Consolidated Standards of Reporting Trials) statement comprises a series of initiatives to improve reporting of randomised clinical trials ~\citep[RCTs, ][]{hopewell2025consort}. Since 2010, \textsc{consort} has recommended that all RCT papers should include the reason for the trial ending~\citep{schulz2010consort}. \cite{moher2012consort}, elaborating on the \textsc{consort} guidelines, explain that this is because ``Readers will likely draw weaker inferences from a trial that was truncated in a data‑driven manner versus one that reports its findings after reaching a goal independent of results.'' \cite{marschner2020sensitivity} note this statement and question its validity, while agreeing that RCT papers should report why the trial ended. The guidelines do not discuss what inferences readers will make from a trial with interim analyses that did not end early for benefit, which tend to underestimate the treatment effect. The widespread citation of \textsc{grade} and \textsc{consort} guidelines underscores their significant influence on research practices; consequently, a rigorous evaluation of their validity—particularly regarding adaptive designs—is essential.

This perceived problem has been addressed in a number of ways. \cite{walter2019randomised} suggest considering each study of a meta-analysis as belonging to one of two subgroups: those with and without stopping rules. They meta-analyse the two subgroups separately and combine the two results with weights approximately equalling the observed probabilities of stopping early and not stopping early, in order to obtain an overall summary estimate that is unbiased. \cite{marschner2020sensitivity} present two courses of action for obtaining unbiased summary estimates: the first is to remove all studies that were permitted to end early. The second is to remove studies that stopped early while adjusting estimates of studies that were permitted to stop early but did not (\textit{nontruncated} studies). Both approaches are found to be inferior to simply including all studies in the meta-analysis.

It has been shown that removing group sequential designs or otherwise reducing their contribution in a meta-analysis does not improve the quality of the meta-analysis. \cite{schou2013meta} investigated the effect on bias, in both common-effect and random-effect meta-analysis, of including group sequential designs that stopped early for benefit. They make a distinction between sequentially monitored studies in general, and the subset of these that stopped early, that is, studies that were \textit{truncated}. Their conclusion is that including truncated trials in a meta-analysis does not cause bias, and that excluding them would introduce ``substantial''  bias. Senn has shown that including trials that allow early stopping for benefit does not cause bias in a common-effect meta-analysis consisting solely of designs of this type, as any overestimation is balanced out by underestimation in adaptive trials that do not stop early~\citep{senn2014meta}. Todd examined the effect on bias of including one or more group sequential designs in a common-effect meta-analysis, and concluded that their inclusion does not cause an increase in bias~\citep{todd1997incorporation}.

In meta-analysis, studies are weighted by the degree of information that they each provide. This concept is intuitive: assigning greater weight to more informative sources of information is natural and uncontroversial in both everyday life and meta-analysis, though at least one paper has questioned this approach~\citep{shuster2010empirical}. A simulation study may be compared conceptually to a meta-analysis in that both are collections of studies for which an effect size estimate is summarised. However, in simulation studies, the effect size estimate and its bias are summarised using the simple mean, without any weighting of the individual studies to reflect their different degrees of information. This is the case even for simulation studies of adaptive designs, where individual study repetitions may provide considerably different degrees of information, for example due to differing sample size. In other words, for adaptive designs, the precision of an estimator of treatment effect can be considered random.

We propose a new measure of bias, \textit{precision-weighted bias}, that accounts for differences in precision in simulation study repetitions. The precision-weighted bias is a weighted mean of the unconditional error of an estimator, weighted by the relative precision of the estimates.

\subsection{Aims}

The aims of this paper are to propose precision-weighted bias as a performance measure that accounts for the differences in information across simulation study repetitions, to explore the properties of this measure, to encourage its use in simulation studies as a complement to conditional and unweighted unconditional bias and to explore its implications for adaptive designs in meta-analysis. Any reference to ``unconditional bias'' or simply ``bias'' should be considered as unweighted unconditional bias.

In the Methods section we begin with a brief motivating example, then derive the precision-weighted bias and describe the scope of our simulation studies. The Results section shows these simulation studies in detail. In the Discussion section, we provide guidance for meta-analysts who must decide how to deal with adaptive design studies and also describe other areas where precision-weighted bias may be useful.

\section{Methods}

\subsection{Motivating example}

Consider a treatment that has a null effect on some continuous outcome, that is, $\theta=0$, with the distribution having variance $\sigma^2$. This treatment effect is measured in a series of identical adaptive design trials that each stop after $N/2$ participants if $\hat{\theta} > 0$, otherwise continue to the maximum number of participants $N$. In other words, the trials have two stages of equal size, with an interim analysis at the end of the first stage, and will stop at this interim analysis if the estimated treatment effect is positive. We stress that this design is for example only, and not useful in practice. We denote as $B$ the conditional bias for trials that stop early. We define precision-weighted bias formally in subsection \ref{sec:pwb} below, but informally, precision-weighted bias in this context is a weighted mean of the biases of a set of studies, where the weights represent the relative amount of information provided by each study. Conditional, unconditional and precision-weighted bias at various points in the trial for different subsets of trial results are shown in Table~\ref{tab:intro_example}. 

\begin{table*}[h]
    \centering
    \caption{Bias for each stage and overall, for a two-stage trial with $\theta=0$, interim analysis at $N/2$ and stopping early if $\hat{\theta} > 0$. Note: precision-weighted bias is unconditional.}
    \label{tab:intro_example}
    \resizebox{\ifdim\width>\linewidth\linewidth\else\width\fi}{!}{
    \begin{tabular}[t]{rcccc} 
        \toprule
        Bias                        & Stage 1 only  & Stage 2 only  & Overall & Precision of overall\\
        \midrule
        Conditional (early stopped) & $B$           & ---           & $B$ & $(N/2)/\sigma^2$\\
        Conditional (Continued)     & $-B$          & 0             & $-B/2$ & $N/\sigma^2$\\
        Unconditional (all)         & ---           & ---           & $\frac{B + (-B/2)}{2} = B/4$ & $((N/2)+N)/2\sigma^2$ \\
        Precision-weighted (all)    & ---           & ---           & $\frac{B + 2(-B/2)}{1+2} = 0$ & $((N/2)+N)/2\sigma^2$   \\
        \bottomrule
    \end{tabular}}
\end{table*}

Placing the stopping boundary at zero, equal to the true treatment effect, means that half of the trials will stop early, after $N/2$ participants (when $\hat{\theta} > 0$ at the interim analysis) and the other half will continue, recruiting a total of $N$ participants (when $\hat{\theta} \leq 0$ at the interim analysis). The conditional bias at stage 1 for trials that continued is $-B$, as it is the additive inverse of the conditional bias for trials that stopped early, defined above as $B$. The overall unconditional bias is the simple mean of the end-of-trial conditional biases, while the precision-weighted bias is a weighted mean. Placing the interim analysis at $N/2$ means that the continued trials have exactly double the precision of the early stopped trials, because they contain double the information of the early stopped trials. For a more general example of weighting studies based on whether they stopped early, see \cite{senn2014meta}.

\subsection{Precision-weighted bias}\label{sec:pwb}

In meta-analysis, the common-effect model assumes that all included studies estimate the same true treatment effect -- they have the same treatment effect in common. Any difference between an observed treatment effect estimate and this true treatment effect is assumed to be the result of random error. Let study $i$ have point estimator $Y_i$ and weight $W_i=1/\text{var}(Y_i)$, which is a measure of the study's precision, or how much information each study contains. In an adaptive design, $W_i$ is primarily driven by the sample size of study $i$ at termination. $W_i$ is often regarded as a fixed value, but here we recognise that it may be a random variable \citep{jackson2018should}, as for example in adaptive trials that may stop early. We recognise also that $W_i$ may be correlated with $Y_i$. We let $\theta$ be the true value of the treatment effect estimated by each $Y_i$; we make no assumption about whether the $Y_i$ are unbiased for $\theta$.

The common-effect summary effect estimator is $\sum W_i Y_i / \sum W_i$ with bias 

\[
\mbox{Bias} = \mathbb{E} \left[ \frac{\sum W_i Y_i}{\sum W_i} \right] - \theta.
\]

Using a Taylor approximation $\mathbb{E}\left[A/B\right] \approx \mathbb{E}[A]/\mathbb{E}[B]$ for small $\text{Var}(B)$ \citep{kendall69}, we get

\begin{equation}\label{eq:bias}
\mbox{Bias} \approx \frac{\mathbb{E} \left[ \sum W_i Y_i \right]}{\mathbb{E} \left[ \sum W_i \right]} - \theta
= \frac{\sum \mathbb{E} \left[ W_i \right] \frac{\mathbb{E} \left[ W_i (Y_i - \theta) \right]}{\mathbb{E} \left[ W_i \right]} }{\sum \mathbb{E} \left[ W_i \right]}.
\end{equation}
which is valid for small $\text{Var}(\sum W_i) / \mathbb{E}\left[ \sum W_i \right]^2$: this occurs for example if there is a large number of independent and identically distributed  studies, or if the coefficient of variation of each $W_i$ is small. We will use simulation studies to explore the sensitivity of our results to this approximation.

Next we define the precision-weighted bias: 
\begin{align}
    \text{PWB} & = \frac{\mathbb{E} \left[ W_i (Y_i - \theta) \right]}{\mathbb{E} \left[ W_i \right]}, \text{ with estimator} \nonumber\\
    \widehat{\text{PWB}} & =  \frac{\sum W_i \left(Y_i - \theta \right) }{ \sum W_i}. \label{eq:pwb}
\end{align}

Equations (\ref{eq:bias}) and (\ref{eq:pwb}) provide our core proposal: \textbf{The bias in a common-effect meta-analysis is approximately a precision-weighted average of the precision-weighted biases in each of the studies}. Directly from this we make two assertions:

\begin{enumerate}
  \item If all studies have zero precision-weighted bias, then the common-effect meta-analysis has approximately zero bias;
  \item If an existing meta-analysis is unbiased and a new study is added, whether bias is introduced depends on whether the new study has zero precision-weighted bias, and not on whether it has zero (unweighted) bias. Furthermore, the magnitude of this introduced bias will be approximately proportional to the relative precision of the added study and to the precision-weighted bias of the added study.
\end{enumerate}

Using Equations (\ref{eq:bias}) and (\ref{eq:pwb}), we can obtain an expression for the change in bias that results from adding a study to a (common-effect) meta-analysis of $K$ studies:

\begin{align}
    \text{Bias}_{inc} & \approx \frac{\sum\limits_{i=1}^{K+1} W_i \text{PWB}_i }{\sum\limits_{i=1}^{K+1} W_i} \nonumber \\ & = \frac{\sum\limits_{i=1}^K W_i \text{PWB}_i + W_{K+1} \text{PWB}_{K+1} }{\sum\limits_{i=1}^{K} W_i + W_{K+1}} \nonumber \\
    & = \frac{\left(\sum\limits_{i=1}^K W_i \right) \text{Bias}_{exc} + W_{K+1} \text{PWB}_{K+1} }{\sum\limits_{i=1}^{K} W_i + W_{K+1}}  \nonumber \\
     \implies \text{Bias}_{inc} - \text{Bias}_{exc} & \approx \frac{W_{K+1} \left( \text{PWB}_{K+1} - \text{Bias}_{exc} \right) }{\sum\limits_{i=1}^{K+1} W_i}. \label{eq:bias_difference}
\end{align}

Equation (\ref{eq:bias_difference}) leads to a further, more general assertion:

\begin{enumerate}[resume]
  \item The change in bias when adding a new study to a common-effect meta-analysis is approximately equal to the weighted difference between the precision-weighted bias of the new study and the bias of the original meta-analysis, where the weight pertains to the weight of the new study with respect to the (updated) meta-analysis.
\end{enumerate}

We will make comparisons of unconditional and precision-weighted bias in the simpler case of single studies (both non-adaptive and adaptive) before setting out tests of the above assertions.

\subsection{Simulation studies: individual studies}

To obtain the precision-weighted bias of a trial design, we undertake a simulation study comprising many repetitions of the trial. We estimate the treatment effect and obtain the error and standard error (SE) for each repetition. The precision-weighted bias is then estimated by taking the weighted mean of the errors, where the weight is the reciprocal of the squared SE.

The non-adaptive design we examine is a single-arm binary outcome trial. Define $\theta_0$ as an ``uninteresting'' response probability typical for standard of care and $\theta_1$ as an ``interesting'' response probability that would make a treatment worth pursuing. The null hypothesis is $H_0: \theta \leq \theta_0$, while the type-I error-rate and power are defined as $P(\text{reject } H_0 \vert \theta=\theta_0)$ and $P(\text{reject } H_0 \vert \theta=\theta_1)$ respectively. The the trial is considered a success and the null hypothesis rejected if a pre-specified number of responses is observed. While sample size is typically chosen to achieve a certain power, here the sample sizes ($N \in \{233, 127\}$) were chosen to match those used for the adaptive design, detailed in the following paragraph.

The adaptive design we examine as a single study is the Simon design: a single-arm, two-stage binary outcome trial design that allows stopping only for futility at the end of the first stage~\cite{simon1989optimal}. In addition to the final or maximum sample size and the number of responses required to reject the null hypothesis, for the Simon design we need to specify the number of participants that comprise each stage and the number of responses required to continue the trial (at the interim analysis). Together, these properties determine the operating characteristics, including the expected number of participants in the trial, known as the expected sample size. We set the required (one-sided) type-I error-rate to be 0.05 for response probability $\theta_0$ and required power to be 90\% for response probability $\theta_1$. We chose the Simon design that minimises the expected sample size when $\theta=\theta_0$, while satisfying the required type-I error-rate and power. For $(\theta_0=0.5, \theta_1=0.6)$, the resulting Simon design has maximum sample size $N=233$ with interim analysis at $n_1=104$. For $(\theta_0=0.8, \theta_1=0.9)$, the resulting Simon design has maximum sample size $N=127$ with interim analysis at $n_1=44$. All computation and simulation was done using \textsc{R} software, version 4.5.3, with Simon designs obtained using the R package \texttt{clinfun}~\citep{R,clinfun}.

\textbf{Aims:} Examine precision-weighted bias and unconditional bias for both non-adaptive and adaptive designs for a binary outcome single-arm trial using an unadjusted estimator.

\textbf{Data generating mechanisms:} We simulate binary data for two sample sizes $N \in \{233, 127\}$, across true response probabilities  $\theta \in \{0.1, 0.2, \dotsc, 0.9\}$. The same simulated data are used to evaluate both the non-adaptive and the adaptive designs.

\textbf{Estimands:} Our estimand $\theta$ is the response probability.

\textbf{Method of analysis:} The response probability is calculated directly as the proportion of responses and its SE.

\textbf{Performance measures:} Unconditional bias, precision-weighted bias.
 
\textbf{Number of repetitions:} $n_{sim}=100,000$.

\subsection{Simulation studies: Meta-analyses}

The following simulation studies each add an adaptive design study to a common-effect meta-analysis containing only non-adaptive designs. The purpose of these simulation studies is to test our core proposal, that the bias in a common-effect meta-analysis is approximately a precision-weighted average of the precision-weighted biases in each of the studies, specifically with regard to our assertions above.

\textbf{Aims:} 
\begin{itemize}
    \item To quantify the bias of a common-effect meta-analysis comprised solely of studies with zero precision-weighted bias.
    \item To compare the change in bias when adding a new study to a common-effect meta-analysis to the weighted difference between the precision-weighted bias of the new study and the bias of the original meta-analysis (where the weight pertains to the weight of the new study with respect to the (updated) meta-analysis).
\end{itemize}

\textbf{Data generating mechanisms:} 
For two true response probabilities $\theta \in \{0.5, 0.3\}$, we simulate binary data for $K \in \{10, 100\}$ studies: one Simon design (using the same design as that used for $(\theta_0=0.5, \theta_1=0.6)$ in the individual studies above) with $N=233$ and all others non-adaptive, single-arm studies with $N=233$, to match the Simon design. For each combination of $\theta$ and $K$, we form a meta-analysis, each consisting of one adaptive and $K-1$ non-adaptive studies.

\textbf{Estimands:} Our estimand $\theta$ is the summary response probability, for each meta-analysis.

\textbf{Method of analysis:} The response probability is estimated for each meta-analysis as the weighted mean of each study's point estimate $Y_i$, where the weight of each study $W_i$ is the inverse of its squared SE.

\textbf{Performance measures:} 

\begin{itemize}
    \item Unconditional and precision-weighted bias of each study.
    \item Unconditional and precision-weighted bias in each meta-analysis, before and after the addition of the corresponding Simon study.
    \item The weight of the added Simon study with respect to the updated meta-analysis.
\end{itemize}

\textbf{Number of repetitions:} $n_{sim}=100,000$.

\section{Results}
\subsection{Simulation studies: individual studies}

\begin{figure*}
    \centering
    \includegraphics[width=0.75\linewidth]{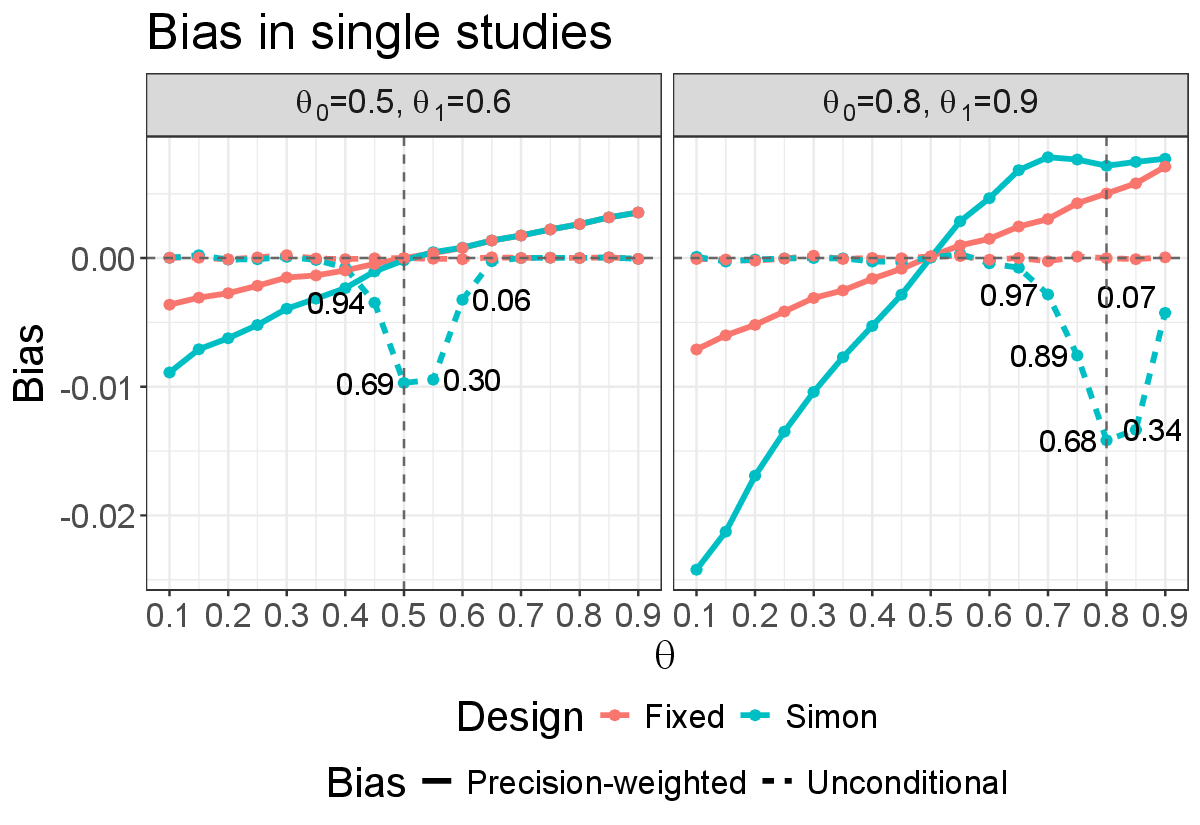}
    \caption{Bias in simulated fixed design and Simon design trials with ${(\theta_0=0.5, \theta_1=0.6)}, N=233$ (left) and ${(\theta_0=0.8, \theta_1=0.9)}, N=127$ (right). Annotation is proportion of trials stopped early (selected).}
    \label{fig:single}
\end{figure*}

Figure~\ref{fig:single} compares the unconditional and precision-weighted bias in the fixed and Simon design simulation studies. Where the proportion of early-stopped trials is in the interval $\left(0.01, 0.99\right)$, this value is shown on the figure.

Alt text: Two line graphs showing how bias -- both unweighted unconditional and precision-weighted -- changes with response probability theta, for the Simon design and the non-adaptive design, under two different sets of design characteristics. In both, the unconditional bias for the Simon design shows a ``dip``, that is, bias is negative, around the theta nought.

For the fixed, non-adaptive design (red lines), the unconditional bias (dashed) is close to zero for all $\theta$. The precision-weighted bias (solid) increases linearly with response probability $\theta$, passing approximately through the origin when $\theta=0.5$. The magnitude of the precision-weighted bias is greater for the smaller trial ($N=127$, right) than the larger trial ($N=233$, left).  Monte Carlo standard error (MCSE) of bias is also recorded~\citep[defined as $MCSE = \sqrt{\frac{1}{n_{sim}(n_{sim}-1)} \sum_{i=1}^{n_{sim}} (\hat{\theta_i} - \Bar{\theta_i})^2 }$,][]{Morris2019May}, for each combination of design type and $\theta$. The MCSE quantifies the uncertainty of the results due to using simulation. The MCSE was less than $0.0003$ in each case.

For the Simon design (blue lines), the unconditional bias (dashed) is low when almost all repetitions have the same realised sample size, i.e., the probability of early stopping is close to zero or one. The magnitude of the bias increases as the probability of early stopping for lack of benefit (i.e. probability of adaptation) approaches 0.5, which is expected~\citep{Whitehead1986Dec,robertson2021point,marschner2021general}, and occurs when response probability $\theta=\theta_0$, the response probability under the null hypothesis. The precision-weighted bias (solid) generally increases with $\theta$ and is negative for $\theta < 0.5$ and positive for $\theta > 0.5$, as for the fixed design. The magnitude of the unconditional bias is typically greater than that of the precision-weighted bias when the probability of early stopping is  between 0.01 and 0.99.

\begin{figure}[h!]
    \centering
    \includegraphics[width=1\linewidth]{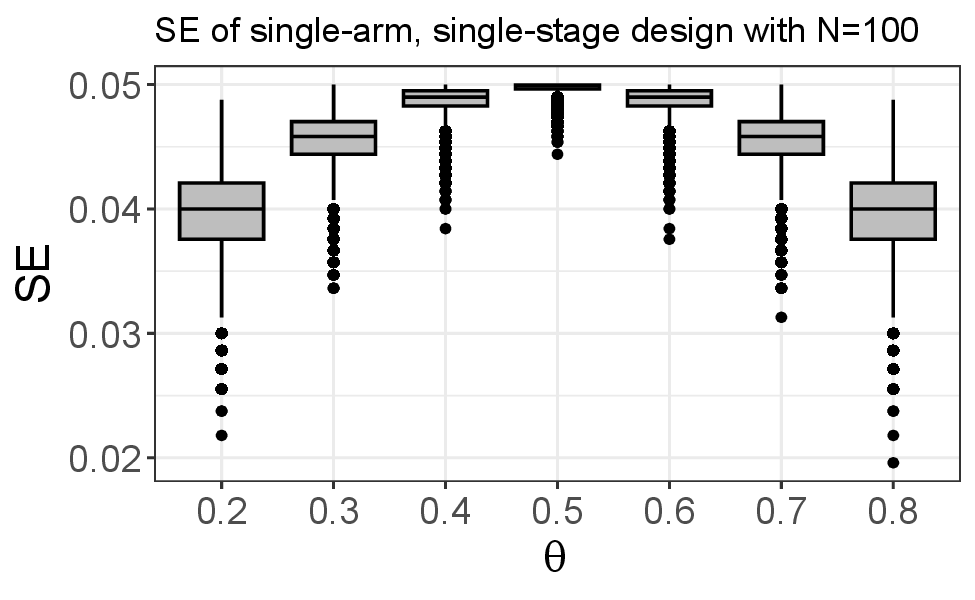}
    \caption{Response probability SE for each simulation.}
    \label{fig:non-AD_SE}
\end{figure}

The increased magnitude of precision-weighted bias can be explained by considering the SE of the response probability estimates. The precision-weighted bias is a function of the SE of the response probability estimate, with simulation study repetitions with smaller SE given greater weight than repetitions with greater SE. When $\theta$ is close to 0.5, there is low variation in the distribution of SEs, meaning that weights are (approximately) equal across simulation study repetitions. Conversely, when $\theta$ is closer to zero or one, there is greater variation in SE and furthermore, the distribution is skewed, with a small number of extremely small SEs (Figure~\ref{fig:non-AD_SE}). 

Alt text: a series of boxplots showing the distribution of SE for different values of theta. The plot overall is symmetrical, with boxplots further from theta=0.5 having a wide distribution, narrowing as theta approaches 0.5, at which point the ``box'' itself resembles a line.

These extremely small SEs are produced by simulation study repetitions where $\hat{\theta}$ is very close to zero or one. When calculating the precision-weighted bias, these repetitions, which have greater (absolute) bias than those in which $\hat{\theta}$ is less extreme, are given the greatest weight. Thus, there is dependence between the estimate $\hat{\theta}$ and the corresponding SE when the response probability estimate is close to zero or one. This phenomenon has been described previously by~\cite{schmid2020handbook}, and occurs not only for simulation studies of individual non-adaptive trials, but also simulation studies of meta-analyses of non-adaptive trials. In parallel to this issue is the more general fact that SE is greater in smaller sample sizes, with the consequence that greater precision-weighted bias is observed in the right-hand subfigure of Figure~\ref{fig:single} (vs the left) for both designs; in the Simon design (vs the fixed design) when $\theta < \theta_0$ and in the Simon design when $\theta < \theta_0$ (vs $\theta \geq \theta_0$).

\subsection{Simulation studies: meta-analysis}

The bias for each individual study comprising the $K=10$ meta-analyses is shown in Table~\ref{tab:individual.bias}, where the results are multiplied by $100$ to allow interpretation of bias as a percentage. MCSE of bias is also reported. Differences in bias across the studies is on the same order of magnitude as the MCSE, indicating that such differences are due to Monte Carlo error.

\begin{table*}[h]
\centering
\caption{\label{tab:individual.bias}Bias, MCSE (both multiplied by 100) and mean sample size for individual studies, meta-analyses of K=10 studies.}
\centering
\resizebox{\ifdim\width>\linewidth\linewidth\else\width\fi}{!}{
\begin{tabular}[t]{rrrrr}
\toprule
  & Bias & PWB & MCSE & Mean N\\
\midrule
\addlinespace[0.3em]
\multicolumn{5}{l}{\boldsymbol{$\theta=0.5$}}\\
\hspace{1em}Study 1 & 0.016 & 0.016 & 0.010 & 233\\
\hspace{1em}Study 2 & -0.008 & -0.008 & 0.010 & 233\\
\hspace{1em}Study 3 & -0.013 & -0.013 & 0.010 & 233\\
\hspace{1em}Study 4 & 0.007 & 0.007 & 0.010 & 233\\
\hspace{1em}Study 5 & -0.017 & -0.018 & 0.010 & 233\\
\hspace{1em}Study 6 & 0.002 & 0.002 & 0.010 & 233\\
\hspace{1em}Study 7 & 0.011 & 0.011 & 0.010 & 233\\
\hspace{1em}Study 8 & 0.000 & 0.000 & 0.010 & 233\\
\hspace{1em}Study 9 & 0.001 & 0.001 & 0.010 & 233\\
\hspace{1em}Mean 1-9 & 0.000 & 0.000 & NA & NA\\
\hspace{1em}Simon & -0.950 & 0.000 & 0.012 & 145\\
\addlinespace[0.3em]
\multicolumn{5}{l}{\boldsymbol{$\theta=0.3$}}\\
\hspace{1em}Study 1 & 0.013 & -0.162 & 0.009 & 233\\
\hspace{1em}Study 2 & 0.007 & -0.169 & 0.010 & 233\\
\hspace{1em}Study 3 & -0.009 & -0.185 & 0.010 & 233\\
\hspace{1em}Study 4 & 0.004 & -0.170 & 0.009 & 233\\
\hspace{1em}Study 5 & 0.010 & -0.165 & 0.009 & 233\\
\hspace{1em}Study 6 & 0.006 & -0.169 & 0.009 & 233\\
\hspace{1em}Study 7 & -0.014 & -0.189 & 0.009 & 233\\
\hspace{1em}Study 8 & 0.007 & -0.168 & 0.009 & 233\\
\hspace{1em}Study 9 & -0.007 & -0.183 & 0.010 & 233\\
\hspace{1em}Mean 1-9 & 0.002 & -0.173 & NA & NA\\
\hspace{1em}Simon & 0.005 & -0.402 & 0.014 & 104\\
\bottomrule
\end{tabular}}
\end{table*}

For $\theta=0.5$, the Simon design has greater unconditional bias ($-0.950$) than precision-weighted bias ($0.000$). For $\theta=0.3$, the non-adaptive designs have some precision-weighted bias ($-0.173$) compared to both the unconditional bias ($0.002$) and the precision-weighted bias in the $\theta=0.5$ case ($0.000$). This is examined and explained in the preceding simulation studies of single designs, above.

Table \ref{tab:summary.bias} presents bias and MCSE  for the summary estimates of the meta-analyses (excluding and including the Simon design). It also reports the change in bias due to including the Simon design, and the bias, precision and weights of the Simon designs.

\begin{table*}[h]
\centering
\caption{\label{tab:summary.bias}Detailed bias and MCSE results for meta-analysis (multiplied by 100), where $W_S$ is the weight of the Simon design in the meta-analysis given that the sum of the weights equals 1, i.e., $\sum W_i=1$. ``Precision'' is the mean precision of Simon design as a factor of the overall mean precision of remaining studies. We define $\Delta$ as the difference in bias of the meta-analyses (including minus excluding the study).}
\centering
\resizebox{\ifdim\width>\linewidth\linewidth\else\width\fi}{!}{
\begin{tabular}[t]{rrrrrrrrrrr}
\toprule
\multicolumn{2}{c}{ } & \multicolumn{6}{c}{Meta-analysis} & \multicolumn{3}{c}{Simon} \\
\cmidrule(l{3pt}r{3pt}){3-8} \cmidrule(l{3pt}r{3pt}){9-11}
\multicolumn{2}{c}{ } & \multicolumn{3}{c}{Excluding} & \multicolumn{2}{c}{Including} & \multicolumn{1}{c}{ } & \multicolumn{1}{c}{ } & \multicolumn{1}{c}{ } & \multicolumn{1}{c}{$W_S(\text{PWB}_{S} - \text{Bias}_{exc})$} \\
\cmidrule(l{3pt}r{3pt}){3-5} \cmidrule(l{3pt}r{3pt}){6-7} \cmidrule(l{3pt}r{3pt}){11-11}
$K$ & \boldsymbol{$\theta$} & $\overline{PWB_i}$ & Bias & MCSE & Bias & MCSE & $\Delta$ & $W_S$ & $\text{PWB}_{S}$ & $\text{Predicted }\Delta$\\
\midrule
10 & 0.5 & 0.000 & 0.000 & 0.003 & -0.005 & 0.003 & -0.005 & 0.071 & 0.000 & 0.000\\
10 & 0.3 & -0.173 & -0.153 & 0.003 & -0.165 & 0.003 & -0.012 & 0.047 & -0.402 & -0.012\\
100 & 0.5 & 0.000 & 0.000 & 0.001 & 0.000 & 0.001 & 0.000 & 0.007 & -0.023 & 0.000\\
100 & 0.3 & -0.174 & -0.173 & 0.001 & -0.173 & 0.001 & -0.001 & 0.004 & -0.389 & -0.001\\
\bottomrule
\end{tabular}}
\end{table*}

\subsubsection{Test of core proposal and assertion (1)}
Our core proposal is that the bias in a common-effect meta-analysis is approximately a precision-weighted average of the precision-weighted biases in each of the studies. These values are shown individually for $K=10$ in Table~\ref{tab:individual.bias} (columns \textit{Bias} and \textit{PWB}), and as means in Table~\ref{tab:summary.bias} (columns $\overline{PWB_i}$ and \textit{Bias}. The values are equal to within MCSE for both $K=100$ cases ($0.000$ vs $0.000$ for $\theta=0.5$, $-0.174$ vs $-0.173$ for $\theta=0.3$) and for $K=10, \theta=0.5$ (0.000 vs 0.000), but not $K=10, \theta=0.3$ ($-0.173$ vs $-0.153$). This final result is possibly due to a violation of the assumption of Equation (\ref{eq:bias}), as a result of using a smaller number of studies. Furthermore, the mean precision-weighted bias of the studies gives a better indication of the bias of the meta-analysis than the mean of the unconditional bias: in the $\theta=0.3$ cases, the meta-analyses have considerable bias while the mean unconditional bias (Table~\ref{tab:individual.bias}) is negligible.

The results of these simulation studies also agree with our first assertion, that if all studies in a meta-analysis have zero precision-weighted bias, then the common-effect meta-analysis has approximately zero bias.  

\subsubsection{Test of assertions (2) and (3)}
Our second assertion is that the bias introduced by adding a study to an unbiased common-effect meta-analysis depends on the presence of precision-weighted bias, rather than (unweighted) bias, in the added study. This is tested in the $\theta=0.5$ cases, where the Simon design with considerable bias ($-0.950$ for $K=10$, Table~\ref{tab:individual.bias}) is added to an unbiased meta-analyses. The change in bias is shown in column ``$\mathit{\Delta}$'' of Table 3 (rows 1 and 3). When added, the bias introduced is negligible (-0.005 for $K=10$, $0.000$ for $K=100$).

In assertion (3), we quantify this change in bias to be approximately equal to the weighted difference between the precision-weighted bias of the new study and the bias of the original meta-analysis, where the weight pertains to the weight of the new study with respect to the (updated) meta-analysis. Both the observed and predicted changes are shown in Table~\ref{tab:summary.bias} ($\mathit{\Delta}$ and \textit{Predicted} $\mathit{\Delta}$ respectively). To three decimal places, the observed change in bias is equal to the predicted change for both $K=100$ cases ($0.000$ for $\theta=0.5$, $-0.001$ for $\theta=0.3$) and for $K=10, \theta=0.3$ ($-0.012$), but not $K=10, \theta=0.5$ (actual $-0.005$ vs predicted $0.000$).

\section{Discussion}

Bias is an important consideration in clinical trials. In particular, one may be concerned that bias may be unduly increased in a common-effect meta-analysis when adaptive design studies are included. We have developed a new measurement for examining the bias in the estimated treatment effect of a trial design, precision-weighted bias, which can be used to quantify the approximate change in bias that would result from including such a trial in a meta-analysis. While we have presented a single precision-weighted bias measure, for use with common-effect meta-analysis, other measures for precision-weighted bias may be developed, for example for use with random-effects meta-analysis.

When considering if including an adaptive study in a meta-analysis will result in an increase in bias, the precision-weighted bias of the study design is a better indicator than the unweighted unconditional bias. Including an adaptive design with unweighted unconditional bias may result in a negligible increase in the bias of the meta-analysis.

We have also shown that in a meta-analysis of single-arm binary outcome studies, the precision-weighted bias of the constituent studies can be a better indicator of overall bias than the unweighted unconditional bias.

The precision-weighted bias of a design is a function of the SE of the response probability estimate. Consequently, in situations where the SE of a design has high variability and/or a strong dependence on the point estimate, such as a single-arm binary outcome design where the response probability is close to zero or one, the precision-weighted bias of the design may be of even greater value. This is the case for both adaptive and non-adaptive designs. In contrast, precision-weighted bias would be lower in, for example, a two-arm randomised controlled trial.

Further aspects of precision-weighted bias may be worth examining in detail. These include exploring precision-weighted bias in continuous outcome trial designs; considering the impact of using adjusted estimators; examining the effect in two-arm designs and considering applying precision weighting to other measures, such as the mean squared error. We acknowledge that we have not explored a variety of scales, such as logit or arcsine.

In summary, the precision-weighted bias is a potentially valuable measurement and one should consider reporting it in both meta-analysis and in simulation studies of adaptive designs.

\section*{Data availability statement}
All data used in this article are simulated, therefore there are no data to be made available.

\section*{Supplementary material}
The code used to simulate and analyse the data are available at the Open Science Framework, \url{https://osf.io/2cp6z}.

\section*{Acknowledgements}
IRW and BCO were supported by the Medical Research Council Programme MC\_UU\_00004/09 and grant UKRI934 from the MRC in partnership with NIHR. ML, DSR, SSV, TJ received funding from the UK Medical Research Council (MC\_UU\_00040/03).

\nocite{*}
\bibliographystyle{biom}
\bibliography{references}

\label{lastpage}

\end{document}